\documentclass[11pt,a4paper]{article}
\usepackage{jstylemarg}
\usepackage[utf8]{inputenc}

\newcommand{\bi}{\begin{itemize}}
\newcommand{\ei}{\end{itemize}}
\newcommand{\bea}{\begin{align}}
\newcommand{\eea}{\end{align}}
\newcommand{\be}{\begin{equation}}
\newcommand{\ee}{\end{equation}}

\makeatletter
\renewcommand*\env@matrix[1][\arraystretch]{%
  \edef\arraystretch{#1}%
  \hskip -\arraycolsep
  \let\@ifnextchar\new@ifnextchar
  \array{*\c@MaxMatrixCols c}}
\makeatother

\author{\quad Charlotte SLEIGHT}

\author{\quad Massimo TARONNA\footnote{Postdoctoral Researcher of the Fund for Scientific Research-FNRS Belgium.}}

\affiliation{Universit\'e Libre de Bruxelles
and International Solvay Institutes\\
ULB-Campus Plaine CP231, 1050 Brussels, Belgium\\}

\affiliation{Department of Physics, Princeton University, Princeton, NJ 08544}

\emailAdd{charlotte.sleight@gmail.com, taronnam@gmail.com}

\title{\centering
\LARGE{Higher-Spin Gauge Theories and Bulk Locality}}

\abstract{We present a no-go result on consistent Noether interactions among higher-spin gauge fields on anti-de Sitter space-times. We show that there is a non-local obstruction at the classical level to consistent interacting field theory descriptions of massless higher-spin particles that are described in the free limit by the free Fronsdal action, under the assumption that such theories arise from the gauging of a global higher-spin symmetry. Our result suggests that the Fronsdal programme for introducing interactions among higher-spin gauge fields cannot be completed without introducing new guiding principles, which could lie beyond the framework of classical field theory.}

\begin{document}

\maketitle

\section{Introduction}
\label{sec::intro}

In this Letter we consider the problem of constructing consistent interacting higher-spin (HS) gauge theories on AdS space-times. By now, the free propagation of massless HS particles is rather well understood \cite{Majorana:1968zz,10.2307/96758,Fierz:1939zz,Fierz:1939ix,Wigner:1939cj,Fronsdal:1978rb,Fang:1978wz,PhysRevLett.58.531,Labastida:1987kw,Metsaev:1995re,Buchbinder:2001bs,Campoleoni:2009gs,Alkalaev:2003qv,Campoleoni:2012th}, but the question of whether mass-less HS particles can interact in a consistent manner is a highly non-trivial one. This question has been the subject of decades of intense efforts, which has over the years gathered increased attention owing (in-part) to the tantalising hypothesis that a HS symmetry may govern the high energy regime of a UV-complete theory of gravity \cite{Fradkin:1986ka}. That this is a possibility within String Theory was argued by Gross in the 80s \cite{Gross:1988ue}, based on the high energy behaviour of string amplitudes \cite{Gross:1987kza,Gross:1987ar}.

Among these efforts, in flat space there are various well-known no-go results \cite{Weinberg:1964ew,Coleman:1967ad,Velo:1970ur,Aragone:1979hx,Weinberg:1980kq,Porrati:2008rm,Bekaert:2010hp,Porrati:2010hm,Taronna:2011kt,Taronna:2012gb,Joung:2013nma,Taronna:2017wbx,Roiban:2017iqg} on the interactions of massless HS particles.\footnote{We remind the reader that the constructive role of no-go theorems, which often allow exceptions, is to highlight which routes should not be taken in the search for self-consistent theories.} Interactions on backgrounds of constant non-zero curvature have drawn more attention due to the prospect\footnote{In particular, in AdS the minimal coupling of a spin-$s$ gauge field to gravity has been shown to be compatible with gauge invariance at the level of cubic interactions by allowing the addition of higher-derivative interaction terms. The $s-s-2$ cubic vertex is given by the Fradkin-Vasiliev vertex \cite{Fradkin:1986qy,Fradkin:1987ks}, whose highest derivative term has $2s-2$ derivatives which is accompanied by a tail of lower derivative terms that includes the two-derivative minimal coupling. This vertex is unique \cite{Boulanger:2008tg}. The question of compatibility of quartic and higher-order interactions with gauge invariance is a non-trivial one, owing to the issue of locality that emerges at these orders -- which we shall explain in the following.} of a manifestly Lorentz covariant minimal coupling of HS gauge fields to gravity \cite{Fradkin:1986qy,Fradkin:1987ks} and the existence of a well-defined HS algebra \cite{Fradkin:1986ka,Eastwood:2002su,Vasiliev:2003ev} that is compatible with the latter \cite{Joung:2013nma}.\footnote{It is worth noting here that by giving up manifest Lorenz covariance (e.g. by working on the light cone) there has been some promising results \cite{Bengtsson:1983pd,Metsaev:1991nb,Metsaev:1991mt,Conde:2016izb,Sleight:2016xqq,Ponomarev:2016lrm} regarding the possibility of minimal coupling in flat space, and corresponding HS algebra \cite{Bekaert:2008sa,Sleight:2016xqq}.}  These attractive features of interactions on anti-de Sitter (AdS) spaces were given further weight by the advent of conjectured holographic dualities between HS gauge theories on AdS and free conformal field theories \cite{HaggiMani:2000ru,WittenTalk,Sezgin:2002rt},\footnote{For reviews see e.g. \cite{Petkou:2004nu,Bekaert:2012ux,Giombi:2012ms,Giombi:2016ejx,Sleight:2017krf}.} and Vasiliev's system \cite{Vasiliev:1990en,Vasiliev:2003ev} (see e.g. \cite{Bekaert:2005vh,Rahman:2015pzl} for reviews) which puts forward a field-theoretical framework to describe putative consistent (semi-)classical interacting theories of HS gauge fields. The existence of possible AdS analogues of flat space no-go results on HS interactions has been little explored to date, with the exception of an extension of the Coleman-Mandula theorem \cite{Maldacena:2011jn}.

A time-honoured approach to constructing interacting theories with a gauge symmetry at the classical level is the Noether procedure \cite{Berends:1984rq}.\footnote{Which can be formulated as a BRST cohomolgical problem \cite{Barnich:1993vg,Stasheff:1997fe,Henneaux:1997bm}.} To this end, given the free theory, one attempts to systematically construct interactions as deformations of the free action/equations of motion through the requirement of gauge-invariance. An important consistency requirement in using such approaches in a field theory setting is that of locality. This is particularly relevant for HS gauge theories: It is well known that a putative interacting theory of HS gauge fields would require us to allow non-local interactions that are unbounded in their number of derivatives at quartic order.\footnote{\label{foo1}This can be understood from the spin-dependent lower-bounds on the number of derivatives present in on-shell cubic-vertices \cite{Bengtsson:1986kh,Metsaev:2005ar,Joung:2011ww} and the spectrum which is unbounded in spin \cite{Fradkin:1986qy,Metsaev:1991nb,Metsaev:1991mt}. Exceptions to this include: Three-dimensional purely Chern-Simons higher-spin gauge theories \cite{Blencowe:1988gj,Bergshoeff:1989ns,Fradkin:1989xt,Campoleoni:2010zq}, which are two derivative. see also CS descriptions \cite{Manvelyan:2013oua} in five-dimensions and \cite{Engquist:2007kz} in generic odd dimensions.} The crucial subtlety in the search for non-trivially interacting HS field theories is whether or not the functional class of such non-localities renders the theory equivalent to the free theory under field re-definitions.\footnote{Such type of non-localities should not be confused with the non-localities introduced in \cite{Francia:2002aa,Francia:2002pt} to remove the trace constraints on the gauge parameter in Fronsdal's off-shell formulation of free higher-spin gauge fields, which are designed to vanish on-shell.} This point was first clarified by Barnich and Henneaux in \cite{Barnich:1993vg}, where this triviality of interactions was shown to arise as a consequence of placing no restriction on the functional class of non-localities. This was further refined in \cite{Taronna:2011kt,Taronna:2017wbx}, where it was shown that a non-trivially interacting field theory is not feasible with a functional class that allows contact interactions that are as non-local as the total exchange amplitude. We refer to such functional classes as non-local obstructions throughout.\footnote{Let us note that this definition can be regarded as a statement about the regularity of the solution to the Noether procedure. For HS gauge theories in flat space, by allowing non-localities of the same class as the exchange amplitudes one can remove the offending terms at the root of Weinberg theorem \cite{Weinberg:1964ew}. This is equivalent to considering distributional solutions to the Ward identities of the type $\delta(\Box)$ in momentum space (as shown in \cite{Sleight:2016xqq}), which require to allow for non-convergent summations over spin and derivatives. In our language, the presence of a non-local obstruction in HS gauge theories would be equivalent having such a distributional solution to the Noether procedure, which we note is moreover consistent with the AdS Coleman-Mandula theorem \cite{Maldacena:2011jn}. Indeed, note that AdS HS amplitudes in Mellin space, which has been proposed as the AdS analogue of momentum space for flat space scattering amplitudes \cite{Penedones:2010ue}, are given by distributions -- see e.g. \cite{Bekaert:2016ezc,Taronna:2016xrm}.}

A key open question is whether or not there exists a weaker notion of non-locality for HS gauge theories which would lie somewhere between locality in the strict sense (which gives no interacting solutions to the Noether procedure at all) and a non-local obstruction as defined above (which gives infinitely many solutions all equivalent to the free theory, in which case the Noether procedure is not constraining). In this work we present a no-go result on the consistent interactions of HS gauge fields on AdS$_{d+1}$ that are described in the free limit by the free Fronsdal action and which arise from the gauging of a global higher-spin symmetry. We find that it is not possible to have interacting classical field theory descriptions of massless HS particles on AdS backgrounds without a non-local obstruction. In particular, we clarify the degree of non-locality that would be needed to construct consistent quartic couplings, which are shown to be as non-local as the 4pt exchange amplitudes in the theory.\footnote{We note that the same question was also explored in a previous work \cite{Bekaert:2015tva}. In contrast to the present Letter, there it was argued that there does exist weaker notion of non-locality for such HS theories. As a point of clarification, in section 6 of \cite{Bekaert:2015tva} it was however incorrectly argued that there are no contributions from single-particle states in the amplitude of the quartic vertex, the presence of which -- as we shall see later on -- would have been the signal of a non-local obstruction. The main argument of \cite{Bekaert:2015tva} involved studying the single poles of a ``flat space analogue'' to the AdS quartic vertex, obtained by replacing each AdS covariant derivative in the vertex with a partial derivative. The ensuing argument assumed that the single-particle contributions for the AdS vertex and its flat space analogue are the same, but this assumption cannot be made when the AdS vertex is not strictly local. The present Letter clarifies that single-particle contributions were disguised in the AdS quartic vertex of \cite{Bekaert:2015tva} as an infinite sum of local contact (i.e. double-trace) contributions that does not converge uniformly.}

This result suggests that any classical interacting field theory description of massless HS particles on AdS would require new guiding principles to circumvent the triviality problem outlined above. We discuss some possibilities at the end of this Letter. Note that this would also further motivate revisiting the interaction problem for massless HS in flat space, for which there are analogous no-go results \cite{Taronna:2011kt,Taronna:2012gb,Taronna:2017wbx,Roiban:2017iqg} due to the appearance of the same type of non-local obstruction.

 In the final section we make contact with recent observations \cite{Alday:2015ota} on the properties of correlators in the dual CFT picture under crossing.

\section{Noether procedure}

The Noether procedure is a systematic scheme to solve for interacting field theories as deformations of free actions governed by gauge and global symmetries: Given the latter, one postulates the existence of a fully non-linear action and gauge symmetries, which are expanded on a given background in weak fields as
\begin{equation}\label{nlexp}
S=S^{(2)}+\sum_{n>2}S^{(n)}\,, \qquad \delta_{\xi} = \delta^{(0)}_\xi+\sum_{n>0}\delta^{(n)}_{\xi}.
\end{equation}
The notation $(n)$ signifies that the corresponding term is power $n$ in the weak fields. 

The requirement of gauge invariance translates into an infinite set of coupled equations:

\begin{equation}\label{Noether}
\delta^{(n-2)}_\xi S^{(2)}+\left(\sum_{k=3}^{n-1}\delta^{(n-k)}_\xi S^{(k)}\right)+\delta^{(0)}_\xi S^{(n)}=0\,,
\end{equation}
$n =2, 3, 4, ...$, whose most general solution at a given order can be written as
\begin{equation}
  S^{(n)}=S_{h}^{(n)}+S_p^{(n)}\,,\label{noetherSol}
\end{equation}
where $S_h^{(n)}$ solves the homogeneous equation
\begin{equation}
  \delta^{(0)}_{\xi}S_h^{(n)}\;\approx\; 0\,,\label{homosol}
\end{equation}
where $\approx$ means on-shell and $S_p^{(n)}$ is the particular solution to the original equation \eqref{Noether} which contains the information about the lower order solutions.  

A particular solution that has a nice physical interpretation is given by \emph{minus} the exchange amplitudes generated by the lower order couplings. For instance, at quartic order 
\begin{equation}\label{psol}
S^{(4)}_p = -{\cal A}^{(4)} \equiv-\left(\mathcal{A}^{\sf s}+\mathcal{A}^{\sf t}+\mathcal{A}^{\sf u}\right),
\end{equation}
where $\mathcal{A}^{\sf s}$ is the 4pt exchange diagram generated by the cubic couplings in the ${\sf s}$-channel etc. That this solves the quartic ($n=4$) consistency condition \eqref{Noether} can be seen extending the analysis from \cite{Metsaev:1991nb,Sagnotti:2010at,Taronna:2011kt} to prove
\begin{equation}
\delta^{(0)}_{\xi}\left(-{\cal A}^{\left(4\right)}\right) \: \approx \: \delta^{(1)}_{\xi}S^{(3)}.
\end{equation}

A further attractive feature of the above choice of particular solution is that the corresponding homogeneous solution 
\begin{equation}\label{homon}
S^{(n)}_h = {\cal A}^{(n)}+S^{(n)},
\end{equation}
is then directly related to ``scattering-like'' observables of the theory \footnote{The intuition being that it is of the form ``exchange''+``contact''.}. This link is especially significant for theories on AdS$_{d+1}$, where one can draw upon the dual interpretation of such observables as correlation functions of single-trace operators on the $d$-dimensional conformal boundary. We make this relationship more concrete in the following.

\subsection{The homogeneous solution, global symmetries and AdS/CFT}

Further constraints are placed on the homogeneous solution $S^{(n)}_h$ if the gauge symmetry \eqref{nlexp} arises from the gauging of a global symmetry. This imposes\footnote{Recall that the Lie algebra bracket $\left[\!\right[{},{}\left]\!\right]$ is inherited from that of the gauge algebra via
\begin{equation}
\delta^{(0)}_{\left[\!\right[\xi_1,\xi_2\left]\!\right]} = \delta^{(0)}_{\xi_1}\delta^{(1)}_{\xi_2} - \delta^{(0)}_{\xi_2}\delta^{(1)}_{\xi_1}, 
\end{equation}
where $\delta^{(0)}_{\xi}$ and $\delta^{(1)}_{\xi}$ are the first two terms in the expansion \eqref{Noether} of the non-linear gauge transformation. The rigid symmetries also close at the level of $\delta^{(1)}$
\begin{equation}
\delta^{(1)}_{\left[\!\right[{\bar \xi}_1,{\bar \xi}_2\left]\!\right]} = \delta^{(1)}_{{\bar \xi}_1}\delta^{(1)}_{{\bar \xi}_2} - \delta^{(1)}_{{\bar \xi}_2}\delta^{(1)}_{{\bar \xi}_1}\:+\:\left(\text{trivial}\right), 
\end{equation}
meaning that {\small $\delta^{(1)}_{{\bar \xi}}$} provides a representation of the rigid symmetries carried by the field content.}
\begin{equation}\label{homoglob}
\delta^{(1)}_{\bar \xi}S^{(n)}_h \: \approx \: 0,
\end{equation}
where the $\bar \xi$ are the gauge parameters associated to the global symmetries, which are obtained by imposing the Killing equations
\begin{equation}
0 = \left[\,\delta_{\xi} \varphi \, \right]_{\varphi=0} = \delta^{(0)}_{\xi} \varphi,
\end{equation}
with solutions given by Killing tensors $\xi = {\bar \xi}$ (see e.g. \cite{1986JMP....27.2693T,Bekaert:2005ka,Bekaert:2006us,Joung:2014qya}). The constraint \eqref{homoglob} originates from the restriction of the $\left(n+1\right)$-th order consistency condition \eqref{Noether} to global gauge parameters {\small ${\bar \xi}$}:
\begin{align}
0 =  \delta^{\left(1\right)}_{\bar \xi}\left(S^{\left(n\right)}_h+S^{\left(n\right)}_p\right)+\sum^{n-1}_{k=2}\delta^{\left(n+1-k\right)}_{\bar \xi}S^{\left(k\right)} \: \approx \: \delta^{\left(1\right)}_{\bar \xi}S^{\left(n\right)}_h.
\end{align}

For theories on AdS$_{d+1}$ backgrounds, for boundary conditions on the bulk fields compatible with conformal symmetry, the form \eqref{homon} of the homogeneous solution is the generating function of connected $n$-point correlation functions of single-trace operators in the dual CFT$_d$ at large $N$ \cite{Gubser:1998bc,Witten:1998qj}. In particular, the boundary value ${\bar \varphi}$ of the bulk field $\varphi$ sources its dual single-trace operator ${\cal O}$, such that
\begin{equation}
   \langle {\cal O}_1 ...\,{\cal O}_n \rangle_{\text{conn.}}= (-1)^n \frac{\delta}{\delta {\bar \varphi}_n} ... \frac{\delta}{\delta {\bar \varphi}_1}S^{\left(n\right)}_h\left[{\varphi_i\big|_{\partial\text{AdS}}={\bar \varphi}_i}\right]. 
\end{equation}

The construction of consistent interactions on AdS can thus be mapped to the classification of consistent conformal correlators, which serve as the scattering observables of the bulk theory. In particular, in the dual CFT picture the global symmetry constraints \eqref{homoglob} are equivalent to Ward identities
\begin{align}
\delta^{(1)}_{\bar \xi}S^{(n)}_h  \approx 0 \iff 0 = \sum_i \langle {\cal O}_1 ...\, \left[Q_{{\bar \xi}},\,{\cal O}_i\right]...\,{\cal O}_n \rangle, 
\end{align}
where $Q_{{\bar \xi}}$ is the charge associated to the Killing tensor ${\bar \xi}$.

\subsection{Higher-spin gauge theories}

The free Fronsdal action \cite{Fronsdal:1978rb} for spin-$s$ gauge fields serves as a starting point to construct theories of interacting HS gauge fields on AdS$_{d+1}$. The free theory alone encodes the global symmetries that govern the spectrum, which can be extracted from the free theory Noether currents \cite{Joung:2013nma}. 
The HS symmetry algebra closing on totally symmetric gauge fields \cite{Fradkin:1986ka,Eastwood:2002su,Vasiliev:2003ev} is unique in generic dimensions \cite{Boulanger:2013zza}. This constrains the spectrum via the global symmetry requirement 
\begin{equation}
\delta^{\left(1\right)}_{\bar \xi}S^{\left(2\right)} = 0.
\end{equation}
The minimal HS-symmetric spectrum consists of a tower of even spin totally symmetric gauge fields $\varphi_s$, $s = 2, 4, 6, ...\,$ and a parity even scalar $\varphi_0$. This is the so-called type A HS theory \cite{Vasiliev:1990en,Vasiliev:2003ev}, and is the unique HS theory in generic dimensions consisting of only totally symmetric fields. This minimal HS-symmetric spectrum forms a closed sub-sector of any theory of totally symmetric HS gauge fields on AdS$_{d+1}$, which we can therefore restrict to throughout. 

At the interacting level, global HS symmetry fixes all interactions and their couplings completely up to field re-redefinitions. This can be seen by analysing the constraint \eqref{homoglob} for HS Killing tensors ${\bar \xi}$, which forces the homogeneous solution \eqref{homon} to take the unique form of correlators of a free scalar conformal theory on the $d$-dimensional boundary when evaluated on-shell with AdS/CFT boundary conditions \cite{Maldacena:2011jn,Boulanger:2013zza,Alba:2015upa}. In particular, for fixed external legs of spins $s_1$-$s_2$-...\,-$s_n$, we have
\begin{equation}\label{homonrep}
S^{\left(n\right)}_h = \langle {\cal J}_{s_1}...\,{\cal J}_{s_n} \rangle_{\text{conn.}},
\end{equation}
where ${\cal J}_{s_i}$ is the spin $s_i$ single-trace operator of twist $\tau = \Delta = d-2$ on the $d$-dimensional boundary that is sourced by the bulk field $\varphi_{s_i}$. The explicit form of \eqref{homonrep} was first given in \cite{Sleight:2016dba} for general $d$ (see also \cite{Didenko:2012tv,Gelfond:2013xt,Bonezzi:2017vha}).

Combined with the particular solution \eqref{psol}, the complete solution to the Noether procedure with the minimal HS spectrum is thus dictated uniquely up to field re-definitions by global HS symmetry:
\begin{subequations}\label{hsactionfixed}
\begin{align}\label{cubic}
S^{\left(3\right)}_{s_1,s_2,s_3} & = \langle {\cal J}_{s_1}{\cal J}_{s_2}{\cal J}_{s_3}\rangle_{\text{conn.}},\\ \label{nordersol}
S^{\left(n\right)}_{s_1,s_2,...\,,s_n} & = \langle {\cal J}_{s_1}...\,{\cal J}_{s_n} \rangle_{\text{conn.}} - {\cal A}^{\left(n\right)}_{s_1,s_2,...\,,s_n},
\end{align}
\end{subequations}
with $n>3$, where ${\cal A}^{\left(n\right)}_{s_1,s_2,...\,,s_n}$ is the tree-level exchange amplitude with external legs of spins $s_1$-$s_2$-...\,-$s_n$, generated by the lower-order couplings. Note that solution \eqref{hsactionfixed} does not assume the AdS/CFT duality since \eqref{homonrep} holds purely as a consequence of global HS symmetry.

At cubic order, the couplings for any triplet of fixed spins are local (up to field re-definitions):
\begin{align}\label{cubicfreesc}
S^{\left(3\right)}_{s_1,s_2,s_3} = \int_{\text{AdS}} {\cal V}_{s_1,s_2,s_3},
\end{align}
with (schematically) \cite{Sleight:2016dba,Sleight:2016xqq}\footnote{See also \cite{Sleight:2016dba,Francia:2016weg,Sleight:2017cax} for off-shell completions.}
\begin{subequations}
\begin{align}\label{v123}
 {\cal V}_{s_1,s_2,s_3} & = g_{s_1,s_2,s_3} \left[\nabla^{\rho_1}...\nabla^{\rho_{s_3}} \varphi_{\mu_1...\mu_{s_1}} \nabla^{\mu_1}...\nabla^{\mu_{s_1}} \varphi_{\nu_1...\nu_{s_2}}\nabla^{\nu_1}...\nabla^{\nu_{s_2}} \varphi_{\rho_1...\rho_{s_3}}+{\cal O}\left(\Lambda\right) \right],\\ 
\label{g123}
g_{s_1,s_2,s_3} & = \frac{1}{\sqrt{N_{\text{d.o.f.}}}}\frac{\pi ^{\frac{d-3}{4}}2^{\tfrac{3 d-1+s_1+s_2+s_3}{2}}}{ \Gamma (d+s_1+s_2+s_3-3)}.
\end{align}
\end{subequations}
At cubic order there is thus no apparent issue of locality, with the couplings \eqref{cubicfreesc} involving a finite number of derivatives \eqref{v123} with finite coupling constants \eqref{g123}. However, the couplings at quartic and higher-orders generically involve an arbitrary number of derivatives (see footnote \ref{foo1}). 

In the following section, we study the (non)-locality of the quartic interactions \eqref{nordersol}, which is possible with the explicit form of the homogenous solution \eqref{homonrep} with $n=4$ and the 4pt exchange diagrams generated by the local cubic couplings \eqref{v123}.

\section{Locality}

For simplicity we restrict to the quartic self-interaction of the scalar ${\varphi}_0$ (\eqref{nordersol} with $n=4$ and $s_i=0$) \footnote{The same conclusion can be drawn for generic integer $s_i$, and is related via HS symmetry.}:
\begin{equation}\label{quarticloc}
S^{\left(4\right)}_{0,0,0,0} = \langle {\cal J}_{0}{\cal J}_{0}{\cal J}_{0}{\cal J}_{0} \rangle_{\text{conn.}} - {\cal A}^{\left(4\right)}_{0,0,0,0},
\end{equation}
where the homogeneous solution reads explicitly
\begin{multline}\label{homo4rep}
\langle {\cal J}_{0}(y_1){\cal J}_{0}(y_2) {\cal J}_{0}(y_3){\cal J}_{0}(y_4)\rangle_{\text{conn.}} 
= \frac{1}{c} \frac{1}{(y_{12}^2y_{34}^2)^{d-2}}\\ \times \left[u^{\tfrac{d}{2}-1}+\left(\frac{u}{v}\right)^{\tfrac{d}{2}-1}+u^{\tfrac{d}{2}-1}\left(\frac{u}{v}\right)^{\tfrac{d}{2}-1}\right],
\end{multline}
in terms of cross-ratios $u=\frac{y_{12}^2y_{34}^2}{y_{13}^2y_{24}^2}$ and $v=\frac{y_{41}^2y_{23}^2}{y_{13}^2y_{24}^2}$,
and $c$ is proportional to the central charge of the boundary theory. The exchange amplitude is given by
\begin{equation} \label{exchamp4rep}{\cal A}^{\left(4\right)}_{0,0,0,0} = \mathcal{A}^{\sf s}_{0,0,0,0}+\mathcal{A}^{\sf t}_{0,0,0,0}+\mathcal{A}^{\sf u}_{0,0,0,0}\,,
\end{equation}
where e.g. 
\begin{align} \label{pschexch}
\mathcal{A}^{\sf s}_{0,0,0,0} & = \sum_{s \in 2 \mathbb{N}} \mathcal{A}^{\sf s}_{0,0|s|0,0},
\end{align}
with each spin-$s$ exchange $\mathcal{A}^{\sf s}_{0,0|s|0,0}$ is generated by the local $0$-$0$-$s$ cubic coupling \eqref{cubicfreesc}.

To study the locality of \eqref{quarticloc}, we first need to perform the sum \eqref{pschexch} over the exchanged spin. For this it is useful to decompose into conformal blocks. For the spin-$s$ exchange, we have \cite{Bekaert:2014cea,Sleight:2017fpc}
\begin{equation}\label{cpwess}
\mathcal{A}^{\sf s}_{0,0|s|0,0}= {\sf c}^2_{{\cal J}_0{\cal J}_0{\cal J}_s} W^{\sf s}_{0,0|s|0,0}+\text{local contact interactions},
\end{equation}
where $W^{\sf s}_{0,0|s|0,0}$ is the conformal block encoding the contributions in the ${\sf s}$-channel from the exchanged spin-$s$ single-particle state \footnote{This type of contribution is the analogue of the pole in the Mandelstam invariant present in flat space exchange amplitudes, while the local contact terms would be polynomials.}. In the CFT picture this is the contribution induced by the dual single-trace primary operator ${\cal J}_s$, with OPE coefficient ${\sf c}_{{\cal J}_0{\cal J}_0{\cal J}_s}$. Given \eqref{cpwess}, in the view of studying locality of \eqref{quarticloc} it is useful to recall the following standard assumptions of field theory when considering the sum over spin in \eqref{pschexch}:
\begin{description}\label{locass}
\item $\mathbb{A}1$. Infinite summations over derivatives do not generate additional single-particle exchanges, in any channel \footnote{Here we are referring to derivative expansions of quartic (and higher-order) contact interactions. In the context of the exchange diagrams \eqref{cpwess}, 4pt contact interactions also arise. They are local for local cubic vertices, but upon summing over spin \eqref{pschexch} to obtain the full exchange amplitude they become unbounded in derivatives. The assumption $\mathbb{A}1$ prevents this derivative expansion from generating single-particle exchanges on top of those already generated by each individual exchange diagram \eqref{cpwess}.}.
\item $\mathbb{A}2$. Summations over spin do not generate additional single-particle exchanges, in any channel.
\end{description}

While $\mathbb{A}1$ and $\mathbb{A}2$ are not necessary from an S-matrix perspective, in field theory they provide necessary conditions for single-particle exchanges to arise only from cubic graphs at quartic order. I.e. in a field theory setting, in forgoing $\mathbb{A}1$ and $\mathbb{A}2$ one would anyway encounter non-local obstructions as defined in the introduction. 

In the following we are going argue by contradiction to show that $\mathbb{A}1$ and $\mathbb{A}2$ do not hold in any interacting HS gauge field theory on AdS. I.e. we start by assuming that there is no non-local obstruction and then uncover that this is not the case.

In neglecting the local contact terms in \eqref{cpwess}, the sum over spin \eqref{pschexch} is given by the HS (or twist) block \cite{Alday:2016njk,Alday:2016jfr}:
\begin{align}\label{hsblockdef}
{\cal H}^{\sf s}_{\left(0,0|\tau|0,0\right)} &= \sum^{\infty}_{s=0} {\sf c}^2_{{\cal J}_{0}{\cal J}_{0}{\cal J}_{s}}  W^{\sf s}_{0,0|s|0,0}\\
&=\frac{1}{c} \frac{1}{(y_{12}^2y_{34}^2)^{d-2}}\left[u^{\tfrac{d}{2}-1}+\left(\frac{u}{v}\right)^{\tfrac{d}{2}-1}\right].
\end{align}
In other words, we have
\begin{equation}\label{hsbexch}
    {\cal A}^{(4)}_{0,0,0,0}={\cal H}^{\sf s}_{\left(0,0|\tau|0,0\right)}+{\cal H}^{\sf t}_{\left(0,0|\tau|0,0\right)}+{\cal H}^{\sf u}_{\left(0,0|\tau|0,0\right)}+...
\end{equation}
where the $...$ denote terms which, under the the field theory assumptions $\mathbb{A}$, encode only contact contributions - i.e. no single-particle exchanges - in any channel.

In analogy to conformal blocks, HS blocks represent the contribution to a 4pt function from an entire HS multiplet (in a given channel). Accordingly the homogeneous solution \eqref{homo4rep}, which is invariant under global HS symmetry \eqref{homoglob}, can be expressed purely in terms of HS blocks \eqref{hsblockdef} as
\begin{equation}\label{oooouv}
\langle {\cal J}_{0}{\cal J}_{0}{\cal J}_{0}{\cal J}_{0} \rangle_{\text{conn.}} \\
 = \frac{1}{2}\left[{\cal H}^{\sf s}_{\left(0,0|\tau|0,0\right)}+{\cal H}^{\sf t}_{\left(0,0|\tau|0,0\right)}+{\cal H}^{\sf u}_{\left(0,0|\tau|0,0\right)}\right],
\end{equation} 
which can be verified explicitly from \eqref{homo4rep} and \eqref{hsblockdef}.

Combined with the particular solution \eqref{hsbexch}, we find that the non-local part of the quartic self interaction \eqref{quarticloc} is proportional to the total exchange amplitude \eqref{exchamp4rep}:
\begin{align}
S^{(4)}_{0,0,0,0}  = \langle {\cal J}_{0}{\cal J}_{0}{\cal J}_{0}{\cal J}_{0} \rangle_{\text{conn.}}-{\cal A}^{(4)}_{0,0,0,0} = -\frac{1}{2}{\cal A}^{(4)}_{0,0,0,0}+...\,, \label{fqu0000}
\end{align}
thus uncovering a non-local obstruction at quartic order. 

\subsection{The role of crossing symmetry}
\label{subsec::roleofcr}

We note that the non-local obstruction \eqref{fqu0000} can be traced back to the behaviour of the tower of single-trace operators ${\cal J}_s$ under crossing. To see this, it is instructive to consider terms in the correlator \eqref{oooouv} that are independent solutions to the crossing equation, and their microscopic interpretation:

For a 4pt function of scalar operators ${\cal O}$ of dimension $\Delta$,
\begin{equation}
\langle {\cal O}{\cal O}{\cal O}{\cal O} \rangle = \frac{G\left(u,v\right)}{\left(y^2_{12}y^2_{34}\right)^{\Delta}},
\end{equation}
crossing symmetry is the requirement
\begin{equation}\label{crossingf}
f\left(u,v\right) = v^{\Delta}G\left(u,v\right) = u^{\Delta}G\left(v,u\right).
\end{equation}
Equation \eqref{crossingf} is straightforward to solve in free theories \cite{Braun:2003rp,Eden:2010zz,Alday:2010zy}, with general solution of the form \cite{Alday:2015ota}
\begin{equation}\label{fcrosssol}
f\left(u,v\right) = \sum_{i,j} c_{ij}\,u^{\frac{\tau_i}{2}}v^{\frac{\tau_j}{2}}, \qquad c_{ij}=c_{ji},
\end{equation}
which sums over the twists $\tau_i$ of the operators in the theory. The general solution \eqref{fcrosssol} exhibits that, in free theories, pairs of twist-trajectories are mapped into each other under crossing,
\begin{equation}
\text{twist}\;\tau_1 \quad \overset{\text{crossing}}{\leftrightarrow} \quad \text{twist}\;\tau_2.
\end{equation}
For the correlation function \eqref{oooouv}, the following two functions are the independent solutions to crossing \eqref{crossingf}
\begin{subequations}\label{twisttwist}
\begin{align}\label{hshs}
f_{d-2}\left(u,v\right) & = \frac{1}{c} u^{\tfrac{d}{2}-1} v^{\tfrac{d}{2}-1}\\ f_{2(d-2)}\left(u,v\right) & = \frac{1}{c} \left(u^{\tfrac{d}{2}-1}v^{d-2}+v^{\tfrac{d}{2}-1}u^{d-2}\right).\label{hsdt}
\end{align}
\end{subequations}
The solution \eqref{hsdt} originates from the exchange of operators of twist $\tau = d-2$, which are the single-trace operators ${\cal J}_s$ of the HS multiplet, together with the exchange of double-trace operators of twist $\tau = 2\left(d-2\right)$. These two twist-trajectories thus map into each other under crossing
\begin{equation}\label{stdt}
\text{Regge}_{\text{single-trace}} \quad \overset{\text{crossing}}{\leftrightarrow} \quad \text{Regge}_{\text{double-trace}}.
\end{equation}
The mapping of single-trace contributions to double-trace contributions under crossing is quite generic of CFTs in $d>2$ \cite{Fitzpatrick:2012yx,Komargodski:2012ek,Alday:2015ota} (see also \cite{Alday:2007mf,Kaviraj:2015cxa,Kaviraj:2015xsa}), and is characteristic of CFTs with a local bulk dual \cite{Heemskerk:2009pn}.

On the other hand, the solution \eqref{hshs} is self-dual under crossing, with
\begin{equation}\label{sdhshs}
\text{Regge}_{\text{single-trace}} \quad \overset{\text{crossing}}{\leftrightarrow} \quad \text{Regge}_{\text{single-trace}},
\end{equation}
which is typical of CFTs at/around a point of large twist degeneracy \cite{Alday:2013cwa,Alday:2015ota,Alday:2016mxe}, such as the present case of theories with HS symmetry. It is this property that is responsible for the non-local obstruction \eqref{fqu0000}. Indeed, it is straightforward to see that the absence of such a contribution would mean that one instead has 
\begin{equation}\label{localpartsol}
\langle {\cal O}{\cal O}{\cal O}{\cal O} \rangle_{\text{conn.}} = {\cal A}^{(4)} + ...\,,
\end{equation}
precisely cancelling the single-trace contributions in the particular solution \eqref{psol}, and would thus avert the appearance of the non-local obstruction \eqref{fqu0000}. 

\section{Discussion}
Let us briefly (and inexhaustibly) discuss the possibilities for interacting HS gauge theories that we feel deserve further understanding in the light of the results presented in this Letter.

We stress that a common assumption which leads to non-local obstructions of the type \eqref{fqu0000} is the requirement of non-trivial bulk interactions and observables. In particular, our conclusion does not rule out the possibility that HS theories, both in flat and AdS spaces, could be regarded as exotic topological theories with trivial bulk interactions and topological S-matrix-like observables. This possibility has already been discussed in \cite{Porrati:2008rm,Joung:2015eny,Beccaria:2016syk,Sleight:2016xqq}.

From this perspective, let us emphasise that the non-local obstruction \eqref{fqu0000} should, of course, not be considered as an inconsistency of the boundary theory but rather as the statement that the bulk action reproducing the boundary CFT correlators at leading order in $1/N$ is at most an effective action, while the microscopic description leading to such an effective action in the bulk would lie outside of the so called Fronsdal program \cite{Fronsdal:1978rb}. The microscopic description should rather be that of an exotic topological (string-)field theory (see e.g. \cite{Engquist:2005yt} for some ideas in this direction). From this topological viewpoint, cohomologically non-trivial interactions would then live on the boundary of AdS.\footnote{This further suggests that the possibility the non-local obstruction provides for removing all the bulk interaction terms via field re-definitions is still compatible with the presence of non-vanishing boundary terms and therefore with the non-vanishing nature of the on-shell boundary correlators. Such boundary terms should be related to the HS observables originally proposed in \cite{Colombo:2010fu,Colombo:2012jx}. In our discussion, they play the role of HS blocks \eqref{hsblockdef} uniquely fixed by boundary HS symmetry. One should however not attempt to break such HS blocks in terms of off-shell current exchanges and bulk contact interactions, where the obstruction arises.}

Let us also note that, from this perspective, a microscopic bulk definition of HS gauge theories may be possible in terms of properly defined topological string constructions \cite{Bars:2001ag,Engquist:2005yt}. In this setting, one may also attempt a second quantised string description which could provide the additional input to define proper non-local topological string field theory interactions. It is conceivable that the non-triviality of interactions could be restored by the requirement that the corresponding functional space of field re-definitions is globally defined on the underlying string Hilbert space.\footnote{We thank A. Sagnotti for suggesting this possibility to us.} We plan to investigate closely these options in the near future.

Before concluding, we would like to underline a potentially interesting feature of non-local obstructions in HS theories which we feel deserves further exploration (see also \cite{TaronnaLebedevTalk2017}). This observation is based on the fact that there are two possible sources of non-convergence which come from the violation of assumptions $\mathbb{A}1$ and $\mathbb{A}2$: One given by the summation over spin in the exchange amplitude (violation of $\mathbb{A}2$) and the other is given by the derivative expansion of the quartic vertex (violation of $\mathbb{A}1$). The key point is that improvements to the cubic vertices of each individual exchange diagram generate quartic contact interactions (see e.g. \cite{Taronna:2011kt} and the discussion in section 4.5 of \cite{Bekaert:2014cea}) which, upon performing the non-convergent sum over the spin of the exchanged field to obtain the full exchange amplitude, can be chosen to generate any quartic contact interaction with a non-convergent derivative expansion. The non-local obstruction can thus be shifted between the quartic vertex and the exchange amplitude, where in the latter it can be encoded in improvements to cubic vertices by the virtue of the non-convergent sum over spin. This observation allows to for-see the existence of a fine-tuned field frame in which the quartic contact interaction is vanishing ($S^{(4)} \equiv 0$). We emphasise that the non-local obstruction does not disappear in such a field frame, as it is mapped into the divergent sum over spins of the exchange amplitude. It is reasonable to expect that this procedure carries over to higher-order interactions, with the existence of a finely tuned field frame in which all higher-order interactions are re-defined away ($S^{(n>3)}\equiv0$).\footnote{It would be interesting to see if this could be possible whilst also maintaining strictly local cubic vertices (i.e. with strictly local improvements), however we note that this appears not to be the case in open String Field Theory where the number of derivatives in each cubic vertex is unbounded -- see e.g. \cite{Erler:2004hv,Taronna:2016ats}.} We postpone a more in-depth study of such fine-tuned field frames for the future. The study of such fine-tuned field frames could facilitate contact with a possible stringy description of the theory, which could shed light on how to move beyond the assumptions of this Letter.

\begin{acknowledgments}
 We gratefully acknowledge M. Henneaux for very useful discussions, encouragement and comments on the draft. We also thank G. Barnich, X. Bekaert, M. Bianchi, D. Francia, S. Giombi, R. Metsaev, A. Sagnotti, E. Sezgin, P. Sundell and A. Tseytlin for useful discussions. We benefited from the warm hospitality of the University of Rome Tor Vergata during the final stages of preparing the manuscript. The research of M. T. is partially supported by the FNRS grant FC 6369 and the RSF grant 14-42-00047 in association with Lebedev Physical Institute. C. S. holds a Marina Solvay fellowship. The research of both M.T. and C. S. was partially supported by the INFN, and C.S. by ACRI's (Associazione di Fondazioni e di Casse di Risparmio S.p.a.) Young Investigator Training Program as part of GGI workshop ``New Developments in AdS$_3$/CFT$_2$ Holography''.
\end{acknowledgments}

\bibliography{refs}
\bibliographystyle{JHEP}

\end{document}